\documentclass[aps,prl,10pt,final,twocolumn,showpacs]{revtex4}
\usepackage{graphicx}
\usepackage{array}
\usepackage{amsmath}
\usepackage{amssymb}
\usepackage{upgreek}
\usepackage{units}

\makeatletter
\def\@dotsep{4.5}
\makeatother

\newlength{\colwidth}
\newcommand{\at}[1]{{\big|}_{#1}}

\newcommand{\Ai}{\text{Ai}}					
\newcommand{\Bi}{\text{Bi}}					

\newcommand{\CaN}{C\! a}										
\newcommand{\BoN}{B\! o}										
\newcommand{\Te}{\Theta_{e}}																
\newcommand{\Trec}{\Theta_{e,2}}		
\newcommand{\Tadv}{\Theta_{e,1}}		
\newcommand{\Dg}{\Delta\!\gamma}																
\newcommand{\Dgt}{\Delta\!\widetilde{\gamma}}												
\newcommand{\vN}{v_{\scriptscriptstyle \!N}}	
\newcommand{\hN}{h_{\scriptscriptstyle \!N}}	
\newcommand{\heta}{h_{\gamma'=0}}						
\newcommand{\xeta}{x_{\gamma'=0}}						
\newcommand{\hadv}{h_1}											
\newcommand{\sadv}{s_1}											
\newcommand{\srec}{s_2}											
\newcommand{\Aadv}{A_1}											
\newcommand{\Arec}{A_2}											
\newcommand{\xrec}{x_{r}}	
\newcommand{\sinf}{s^{\scriptscriptstyle(\!\infty\!)}}			
\newcommand{\psp}[2]{p'_{\!#1}\!(#2)}		
\newcommand{\xmax}{x_{\scriptscriptstyle max}}	

\newcommand{\Dadv}{drop~1}
\newcommand{\Drec}{drop~2}

\begin{document}

\title{Non-coalescence of sessile drops from different but miscible liquids:\\
Hydrodynamic analysis of the twin drop contour as self stabilizing, traveling wave}
\date{\today}

\author{Stefan Karpitschka}
\author{Hans Riegler}
\affiliation{Max-Planck-Institut f\"ur Kolloid- und Grenzfl\"achenforschung, Potsdam-Golm, Germany}

\begin{abstract}
{Capillarity always favors drop fusion. Nevertheless sessile drops from different but completely miscible liquids often do \emph{not} fuse instantaneously upon contact. Rather, intermediate \emph{non-coalescence} is observed. Two separate drop bodies, connected by a thin liquid neck move over the substrate. Supported by new experimental data a thin film hydrodynamic analysis of this state is presented. Presumably advective and diffusive volume fluxes in the neck region establish a localized and temporarily stable surface tension gradient. This induces a local surface (Marangoni) flow that stabilizes a traveling wave i.e., the observed  moving \emph{twin drop} configuration. The theoretical predictions are in excellent agreement with the experimental findings.}
\end{abstract}


\maketitle


\emph{Introduction.---}Basic physics tells us that two sessile drops will coalesce as soon as they touch each other because of the reduced interfacial area/energy for a single drop. For drops with identical liquids the instantaneous coalescence after contact has been studied  ~\cite{Aarts:PhysRevLett95,Ristenpart:PhysRevLett97}. Yet, little is known about the fusion of drops with different, but completely miscible liquids. Recent studies show that it can be fundamentally different~\cite{Riegler:Langmuir24,Karpitschka:Langmuir26,Borcia:EPJE34}. The drop fusion can be delayed for a long time. After contact, the drop bodies remain separated in a temporary state of \emph{non-coalescence}, connected only through a thin liquid bridge. This ``twin drop configuration'' moves over the surface (Fig. 1). Presumably a Marangoni flow resulting from the surface energy difference of the two liquids causes the non-coalescence~\cite{Riegler:Langmuir24,Karpitschka:Langmuir26}. An adequate hydrodynamic description of the non-coalescence state has not yet been presented. Especially, it has not yet been explained how a temporarily stationary Marangoni flow is established, how this flow stabilizes the non-coalescence, and how this moves the twin drop system. 

Understanding the phenomenon is of general interest because it is counterintuitive, seemingly ignoring capillarity. The interaction of drops of different composition is relevant e.g. to droplet-based microfluidics~\cite{Lai:LabChip10,Li:LabChip11} and especially to Marangoni flows in thin films that are widely applied for surface cleaning (``Marangoni drying'',~\cite{Leenaars:Langmuir6}). Here we present new experimental data and the first analytical hydrodynamic description. We show that the moving twin drop configuration is a special case of temporarily stationary, self-stabilizing thin film waves that are driven by stationary local surface tension gradients. 

\begin{figure}
	\centering\includegraphics[width=.9\colwidth]{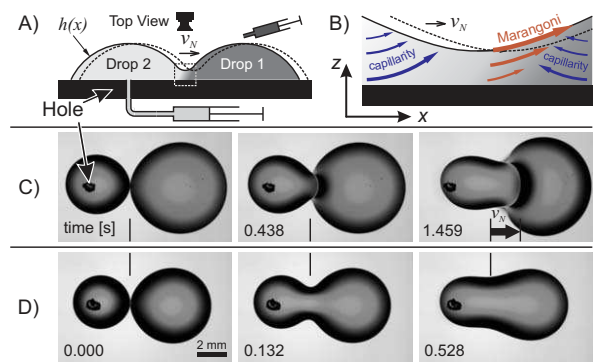}
	\caption{\label{fig:collage}A)~Schematics of the experimental setup/procedure; B)~Flow scheme of the neck region; C)~non-coalescent twin drop movement (different, miscible liquids); D)~instantaneous coalescence (identical liquids).}
\end{figure}

\begin{figure}
	\centering\includegraphics[width=.9\colwidth]{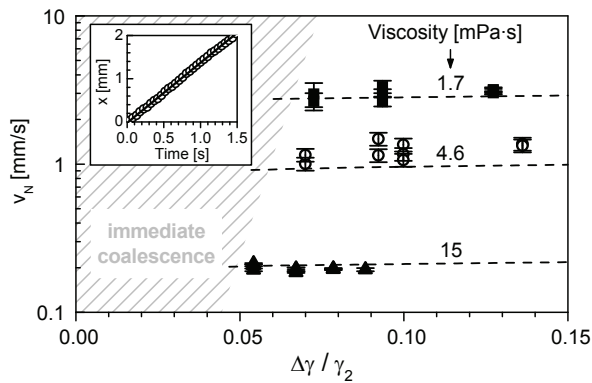}
	\caption{\label{fig:vNeck}Neck displacement velocity $\vN$ as function of the surface energy difference $\Dg/\gamma_2$ (the insert demonstrates the constancy of $\vN$). Dashed lines: analytically calculated $v_N$ (see also Fig.~\ref{fig:cadge}), in quantitative agreement with experimental results.}
\end{figure}

\emph{Experimental Results.---}Figure~\ref{fig:collage} shows A) schematically the experimental setup; B) the neck in detail: capillarity fills the neck from both sides whereas Marangoni sucks out liquid asymmetrically, thus moving the drops with $v_N$ and preventing coalescence~\footnote{See supplemental material for videos, a mechanistic description, an estimation of Taylor-Aris dispersion, and additional information on mathematical details.}; C) non-coalescence; D) instantaneous coalescence. Drop 1 is deposited with a syringe from the top, \Drec~is formed by pumping liquid~2 through a hole in the substrate. More experimental details are found in~\cite{Karpitschka:Langmuir26}. 
With identical liquids, both drops fuse instantaneously upon contact (D). 
With sufficiently different surface tensions (C), the drops do not fuse~\footnote{Liquids: ternary mixtures of 1,2- and 1,3-propanediols and water ($\unit[50]{\%}$ of total mass, viscosity $\eta\approx\unit[4.6]{cP}$). $\Dg$ was tuned by varying relative amounts of diols. Sequence~B: identical liquids (only 1,2-propanediol+water, $\gamma_1=\gamma_2\approx\unit[44.9]{mN/m}$). Sequence~C: \Drec\ as before, \Dadv: $\unit[18.8]{\%}$ 1,2- and $\unit[31.2]{\%}$ 1,3-propanediol; $\Dg\approx\unit[6.1]{mN/m}$. Viscosities were tuned by the water content and matched for each pair of drops.}.  
The main drop bodies ($h''(x)<0$) stay separated in a temporary state of non-coalescence, connected by a thin liquid neck ($h''(x)\gg 0$). 
Through the neck small amounts of liquid continuously flow from \Drec\ (lower surface tension, $\gamma_2$) to \Dadv\ (higher tension, $\gamma_1$). 
The flow slowly reduces the surface energy difference $\Dg=\gamma_1-\gamma_2$. 
Eventually the drops merge (delayed up to minutes~\cite{Karpitschka:Langmuir26}). 
The non-coalescing twin drops move over the substrate with constant velocity, almost independent from $\Dg$ (Fig.~\ref{fig:vNeck}).

\begin{figure}
	\centering\includegraphics[width=.8\colwidth]{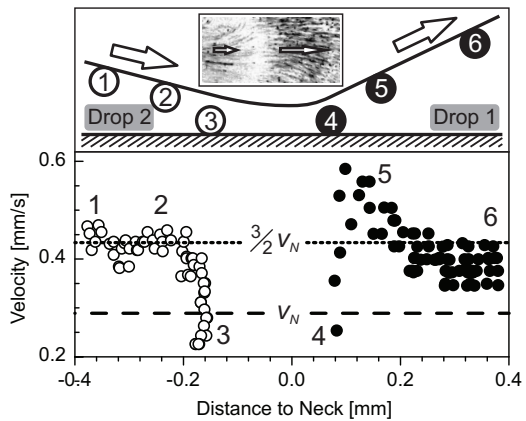}
	\caption{\label{fig:piv2}Flow velocities (substrate frame) of individual microspheres as function of their distance to the neck (moving twin drops, open circles: upstream, \Drec; closed circles: downstream, \Dadv). Dashed line: measured neck velocity $v_N$. Speeds exceeding $\frac{3}{2}v_N$ indicate a (local) surface tension gradient.}
\end{figure}

The flow patterns within moving twin drops were investigated by imaging dispersed fluorescent polystyrene microspheres (Duke Scientific, $d=\unit[1]{\upmu m}$, mass fraction $\approx 2.4\cdot10^{-7}$).  Figure~\ref{fig:piv2} shows their traces in the neck region~[14]. In the substrate frame all microspheres move from left to right. At/near the liquid/air interface of \Drec\ the spheres move with $\approx \frac{3}{2}\vN$ (positions~1 and~2). Close to the neck they touch the substrate ($d\approx$ neck height $h$) and slow down (position~3). On the other side of the neck, in \Dadv, those spheres that where at rest close to the substrate (position 4) get uplifted by the liquid flow through the neck. Spheres which reach the liquid/air interface of \Dadv\ can get accelerated up to $\approx 2\vN$ (position 5). While moving away from the neck region, their speed slows down to $\approx \frac{3}{2}\vN$ (position 6). As discussed below, the speed component in addition to $\frac{3}{2}\vN$ is caused by the local Marangoni flow. 

\emph{Hydrodynamic model.---}The steady state twin drop movement is analyzed by assuming a viscous, laminar Newtonian flow (no slip, no gravity, lubrication approximation). The two drops are approximated by cross sections through two infinitely long cylinders. They are connected by a thin neck and move slowly with constant speed and stationary shape over a planar substrate. They consist of miscible liquids with different surface tensions. Liquid with the lower surface tension continuously flows through the neck into \Dadv. Diffusive-advective liquid mixing establishes a local stationary composition profile because the volume exchange per time is small compared to the overall drop volumes. The composition profile causes locally a surface tension gradient on \Dadv\ close to the neck (for a detailed qualitative description how the Marangoni flow resulting from this gradient stabilizes stationary non-coalescence see [14]).

The (stationary) surface topology $h(x)$ of a liquid film with surface tension $\gamma(x)$, moving on a solid support with capillary number \hbox{$\CaN=\eta\,\vN / \gamma$} is described by the balance between changes of the surface curvature, viscous dissipation and surface energy gradients, respectively~\cite{Oron:RevModPhys69}:
\begin{equation}
	\label{eq:shape}
	h''' = 3\CaN / h^2 - 3\gamma' / (2h\gamma)\text{,}
\end{equation}
Without Marangoni term ($\gamma'=0$), Eq.~\ref{eq:shape} cannot describe a stationary twin-drop profile because $h'''>0$ allows only one inflection point (one drop connected to a neck). With $\gamma'\ne0$ however, a second inflection point respectively a second drop can exist because $h'''$ may change its sign.

The velocity field $u(z)$ within the film is~\cite{Oron:RevModPhys69}:
\begin{equation}
	\label{eq:vfield}
	u =  {z} / {\eta}\big( \gamma' - ({z} /{2}-h)\gamma\,h''' \big)\text{.}
\end{equation}
Eqs.~\ref{eq:shape} and~\ref{eq:vfield} yield the speed at the free surface ($z=h$):
\begin{equation}
	\label{eq:usurf}
	u_{s} =u\at{z=h}=3\vN/2 + h\gamma'/(4\eta)\text{.}
\end{equation}
The Marangoni component increases $u_s$: it sucks liquid out of the neck region and thus favors non-coalescence.

This describes the experimental results (Fig.~\ref{fig:piv2}). For \Drec, the maximum (surface-)velocity is $\approx \frac{3}{2}\vN$ ($=u_s$ with $\gamma'=0$). There is no surface tension gradient on \Drec. At the surface of \Dadv, close to the neck, however $u_{s} > \frac{3}{2}\vN$: there, $\gamma'\ne 0$, obviously caused by liquid~2 flowing through the neck into \Dadv. The data show that $\gamma'\ne 0$ is localized next to the neck ($\approx \unit[0.2]{mm}$, compared to drop sizes $> mm$).

\emph{Localized surface tension gradient.---}
Marangoni forces cause the flow through the neck: they ``pull'' liquid~2 as a thin film onto \Dadv ~where it mixes with liquid~1. This process is approximated by a layer of liquid~2 with the thickness of the neck, $\hN$, spreading on the surface of \Dadv\ with a velocity $\frac{3}{2}\vN$~\footnote{In experiments, the $\gamma'$-contribution to $u_s$ was always significantly smaller than $\frac{3}{2}\vN$; the latter is the zero-order expansion of eq.~\ref{eq:usurf} for small $\gamma'$.}. As liquid~2 flows onto \Dadv\ it is increasingly diluted by diffusion (diffusion constant $D$). Compared to the influx of liquid~2, \Dadv\ is a 
large reservoir of liquid 1. The dilution is a continuous and approximately stationary process. 

With negligible diffusion in flow-direction, this is described by the stationary advection-diffusion equation. In the (moving) contour frame the advection velocity is $\vN/2$:
\begin{equation}
	\label{eq:diffequation}
	D \frac{\partial^2 c}{\partial z^2} = \frac{\vN}{2} \frac{\partial c}{\partial x}\text{,}
\end{equation}
with local compositions~$c\in [0,1]$ from pure liquid~1 to pure liquid~2, respectively. \hbox{$c\at{x=0}=1$} for \hbox{$0<z<2\hN$ and~$0$} elsewhere ($2h$ assures \hbox{$\partial c/\partial z\at{z=\hN}=0$} i.e., non-volatile liquids). $\gamma(x)$ shall depend locally on $c(x,z)$ with $\gamma(x)  = \gamma_2 + \Dg\, c(x,h)$. With this the solution to Eq.~\ref{eq:diffequation} yields: 
\begin{equation}
	\label{eq:dxgamma}
	\gamma' = \frac{\Dg}{2\,\hN}\sqrt{\frac{\BoN}{2\pi}}{\left(\nicefrac{x}{\hN}\right)}^{-\nicefrac{3}{2}}\exp{\left[ -\frac{\BoN}{8}\nicefrac{x}{\hN} \right]}\text{,}
\end{equation}
$\BoN=\vN\hN/D$ is the Bodenstein number, a P\'eclet Number with advection and diffusion  \emph{orthogonal}.

\begin{figure}
	\centering\includegraphics[width=.8\colwidth]{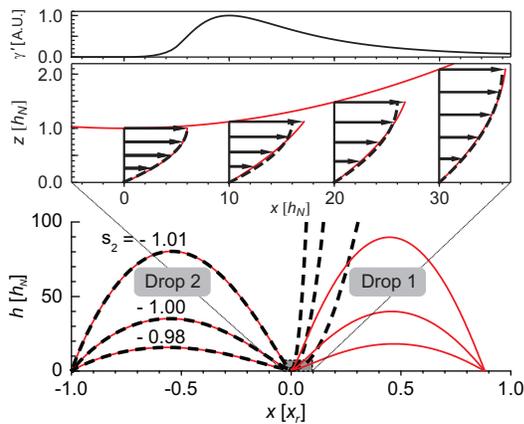}
	\caption{\label{fig:profile}Moving drop contours with different shape parameters $\srec$. Dashed black lines: analytical solutions for homogeneous surface tension (single drop). Solid red lines: non-coalescing twin drop contour (numerical solution) for a localized surface tension gradient $\gamma'$ (top panel), and the corresponding velocity field (middle).}
\end{figure}

$\gamma' _{\scriptscriptstyle max}$ is at $\xmax=\BoN\,\hN/12$ with $\Delta\!x_{\scriptscriptstyle FWHM} \approx 0.225 \BoN\,\hN$ (Fig.~\ref{fig:profile}). With typical experimental parameters ($\vN\approx\unit[1]{mm/s}$, $\hN\approx\unit[10]{\upmu m}$, $D\approx\unit[10^{-10}]{m^2/s}$) $\gamma'$ is $\ne 0$ only close to the neck ($\xmax \approx \unit[0.08]{mm}$, $\Delta\!x_{\scriptscriptstyle FWHM} \approx \unit[0.2]{mm}$). This and typical experimental values of $\Dg\lessapprox0.1\gamma_2$ justifies to assume $\gamma(x) \approx \gamma_2$ whenever $\gamma(x)$ is used explicitly.

\emph{Numerically determined twin drop contour.---}Eq.~\ref{eq:shape} together with Eq.~\ref{eq:dxgamma} is solved numerically. The result (Fig.~\ref{fig:profile}, solid/red lines) is remarkably similar to the experimental findings (explicitly shown in~\cite{Karpitschka:Langmuir26,Borcia:EPJE34}). It indeed reveals a moving twin drop configuration driven by the localized surface tension gradient described by Eq.~\ref{eq:dxgamma}. The contour for $x\leq0$ (\Drec), including in particular $h''\at{x=0}$ is described by Eq.~\ref{eq:shape} with $\gamma'=0$. The main body of \Dadv\ also has a contour very similar to that of \Drec\ i.e., with $\gamma'=0$. This and the limitation of the gradient to the neck section is the key to the following analytical description which aims at predicting $\vN$ from only measurable quantities.

\emph{Analytical approach.---}With the localization of $\gamma'\ne 0$ at the neck region of \Dadv\ the remaining bodies of the two drops can be described by Eq.~\ref{eq:shape} with $\gamma'=0$. Analytic solutions describing a (single) drop contour can be parametrized around a local minimum/maximum ($s=\srec$, $x=0$, and $h=\hN$, see Fig.~\ref{fig:profile}, dashed lines)~\cite{Ford:SIAMrev34,Duffy:ApplMathLett10,Pismen:PhysFluids18}:
\begin{align}
	\label{eq:anax}
	\xeta(s) &= \frac{\hN\left[\Ai(s)\Bi(\srec)-\Ai(\srec)\Bi(s)\right]}{(3\CaN/2)^{\nicefrac{1}{3}}p_{\srec}\!(s)}\text{,}\\
	\label{eq:anah}
	\heta(s) &= \hN/(\pi\,p_{\srec}\!(s))^2\text{,}
\end{align}
with $p_{\srec}\!(s)=\Ai(s)\Bi'(\srec)-\Ai'(\srec)\Bi(s)$. For $\srec\in\;\,[-1.01879,0\,]$ Eqs.~\ref{eq:anax} and~\ref{eq:anah} show a local minimum at $s=\srec$. The abscissa is scaled by the position of the receding contact line $\xrec=\hN(\frac{3}{2}\CaN)^{-\nicefrac{1}{3}}\Ai(\srec)/\Ai'(\srec)$, rendering the plot independent of $\CaN$. The ordinate is scaled by $\hN$ (spherical cap profiles appear distorted). On the right, at $x=0$, the drop contour is connected through a neck of height $\hN$ to the contour of an infinitely large liquid volume: the contour diverges at $x>0$ . At the branch limit $s=\sinf\approx -1.732 + 0.595\,\srec$ (see e.g.~\cite{Duffy:ApplMathLett10} for details), both $\xeta\rightarrow\infty$ and $\heta\rightarrow\infty$.

We assume now that Eqs.~\ref{eq:anax} and~\ref{eq:anah} for $x<0$ individually describe the main body contours of \Drec\ and \Dadv, respectively. For composing the twin drop contour from these two individual contours it is assumed further that the contour of \Drec\ is modified at $ x>0$ (i.e., to the right of the neck in Fig.~\ref{fig:profile}) by the local gradient $\gamma'$ (Eq.~\ref{eq:dxgamma}). In agreement with the experimental and numerical results presented above this modification does not affect the (gradient-free) contour section, describing the main body contour of \Drec\ (Eqs.~\ref{eq:anax} and~\ref{eq:anah} for $x<0$). Instead, the local gradient $\gamma'$ affects the asymptotic curvature of the contour for $x\rightarrow\infty$, far away from the neck region. This gradient-modified curvature is then matched to the apex curvature of another (gradient-free) drop contour, which describes the shape of \Dadv. At the apex of \Dadv, the curvature is always negative whereas the contour of \Drec\ from Eqs.~\ref{eq:anax},\ref{eq:anah} for $ x>0$ (without gradient) always has a positive curvature. Therefore, to match both curvatures, a (local) gradient has to be sufficiently ``strong'' to establish a negative curvature for $x>0$. Then, such matching links the parameters of the individual solutions to a twin drop configuration.   

\emph{Impact of the surface tension gradient.---}The curvature at the neck ($x=0$) is derived from Eqs.~\ref{eq:anax},~\ref{eq:anah}:
\begin{equation}
	\label{eq:neckcurv}
	\heta''(x=0) = -2(3\CaN/2)^{\nicefrac{2}{3}}\srec/\hN\text{.}
\end{equation}
For $\xeta\rightarrow\infty$ (the diverging contour section) the curvature is finite and positive:
\begin{equation}
	\heta''(x\rightarrow\infty) = 2\pi^2(3\CaN/2)^{\nicefrac{2}{3}}\psp{\srec}{\sinf}^2/\hN\text{.}
\end{equation}

With $\gamma'\ne 0$ restricted to the neck region at $x>0$, the curvature change can be approximated: 
\begin{equation}
	\label{eq:limcurv}
	h''_{\infty} \approx \heta''(x\rightarrow\infty) - \frac{3}{2\gamma_2}\int_0^{\infty}\frac{\gamma'}{\heta}dx\text{.}
\end{equation}

Here, the true $h(x)$ is approximated by $\heta$, the contour without gradient. Of course, if we assume a significant impact of the local $\gamma'(x)$, at some distance from the neck, $h$ will be different from $\heta$. However, $\gamma'(x)$ is localized and $h(x)$ grows quickly. Thus, with increasing distance from the neck, $\gamma'/\heta$ (the integrand) rapidly becomes negligible: The integral error remains small, the upper integration limit is irrelevant and can be set to infinity~[14]. Eq.~\ref{eq:limcurv} is integrated analytically by expanding $\heta(x)$ around $x=0$ to second order (see Eq.~\ref{eq:neckcurv}):
\begin{equation}
	\label{eq:mint}
	\frac{3}{2\gamma_2}\int_0^{\infty}\frac{\gamma'}{\heta}dx\approx\frac{3\Dgt}{2\hN} q(k)^{-1}\text{,}
\end{equation}
Here $k=\BoN(\frac{3}{2}\CaN)^{\nicefrac{1}{3}}\sqrt{-\srec}$ and $\Dgt\approx\Dg /\gamma_2$. $q(k)$ is lengthy~[14], but (for typical parameters) can well be approximated in powers of $k$: $q(k) \approx 1 + \sqrt{\pi k}/4 + k(\pi+k/3)/16$. Eqs.~\ref{eq:mint} and~\ref{eq:limcurv} yield:
\begin{equation}
	\label{eq:limcurvsol}
	h''_{\infty} \approx \frac{2}{\hN}\left[{\pi^2(3\CaN/2)}^{\nicefrac{2}{3}}p'_{\srec}\!(\sinf)^2 - \frac{3\Dgt}{4} q(k)^{-1}\right]\text{.}
\end{equation}
The second term indicates that the asymptotic curvature may indeed become negative for sufficiently large $\Dgt$.
 
\emph{Matching the curvatures.---}Eq.~\ref{eq:limcurvsol} estimates the asymptotic curvature for $x\rightarrow\infty$ as a function of a local gradient. As condition for a twin drop configuration this curvature shall match the apex curvature of a gradient-free drop shape (\Dadv). To this end \Dadv\ is described by Eqs.~\ref{eq:anax} and~\ref{eq:anah} (renaming $\srec$ to $\sadv$, $\sadv>0$ to parametrize at the apex (maximum) instead of the neck). The apex curvature is (Eq.~\ref{eq:neckcurv})
\begin{equation}
	\label{eq:apexcurv}
	\heta''\at{apex} = -(3\CaN/2)^{\nicefrac{2}{3}}2\sadv/\hadv\text{,}
\end{equation}
introducing the apex height $\hadv$.

A moving twin drop configuration with a stationary contour means identical capillary numbers $\CaN$ in Eqs.~\ref{eq:limcurvsol} and~\ref{eq:apexcurv}. This yields:
\begin{equation}
	\label{eq:Dgt}
	\Dgt = \nicefrac{4}{3} (3\CaN/2)^{\nicefrac{2}{3}} q(k)
				 \left[\hN \sadv/\hadv + \pi^2 p'_{\srec}\!(\sinf)^2\right]\text{.}
\end{equation}

\begin{figure}
	\centering\includegraphics[width=.9\colwidth]{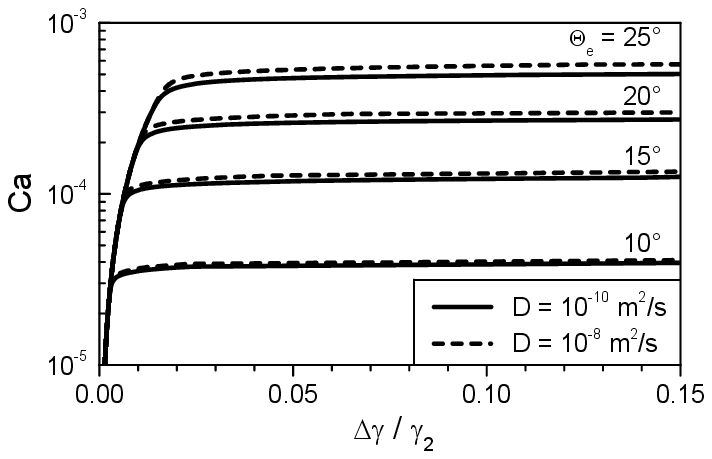}
	\caption{\label{fig:cadge}Capillary number for non-coalescing drops with different contact angles ($\Te$=$\Trec$=$\Tadv$) as function of the surface tension difference $\Dg/\gamma_2$.}
\end{figure}

Equation~\ref{eq:Dgt} still contains the parameters $\srec$, $\sadv$, $\hN$, $\hadv$ (and implicitly $\BoN$). These are linked to experimental parameters such as the \emph{static} equilibrium contact angles $\Tadv$ and $\Trec$, the drop volumes (= areas $\Aadv$ and $\Arec$ under the drop contours ) and $D$. The contour area $\Arec$ ($x<0$) is connected to $\hN$ by:
\begin{equation}
	\label{eq:A2}
	\Arec = (3\CaN/2)^{-\nicefrac{1}{3}}\frac{\hN^2}{\pi^4}\int_{\srec}^{\infty} p_{\srec}\!(s)^{-4}ds\text{,}
\end{equation}

Matching the receding contact line to a microscopic solution~\cite{Eggers:PhysRevLett93,Pismen:PhysFluids18} gives $\srec$ as a function of $\CaN$, with the $\Trec$ and a slip length $\lambda_c$ (besides $\Arec$) as parameters:
\begin{equation}
	\label{eq:s0match}
	\Ai'(\srec)=\frac{\Trec\hN}{6\pi\lambda_c}(3\CaN/2)^{-\nicefrac{1}{3}}\exp\left[-\frac{\Trec^3}{9\CaN}\right]\text{.}
\end{equation}

Matching the advancing contact line of \Dadv\ to a microscopic solution (with the inflection point of solutions Eqs.~\ref{eq:anax} and~\ref{eq:anah} approximated by $s_i\approx s_1-1$) yields~\cite{Pismen:PhysFluids18}:
\begin{equation}
	\label{eq:s1match}
	\Tadv \approx -(12\CaN)^{\nicefrac{1}{3}}\sqrt{s_1\!-\!1}
				 \tanh\!\frac{2}{3}\left[ (s_1\!-\!1)^{\nicefrac{3}{2}} - s_1^{\nicefrac{3}{2}} \right]\text{.}
\end{equation}
$\Aadv$ for \Dadv\ following Eq.~\ref{eq:A2} (with suitable integration limits) determines the apex height $\hadv$.

Now the system is closed. The $\Dgt$ that propels the drops with $\CaN$ results from inserting solutions to Eqs.~\ref{eq:A2},~\ref{eq:s0match}, and~\ref{eq:s1match} into Eq.~\ref{eq:Dgt}. Figure~\ref{fig:cadge} presents the ensuing $\CaN$ as function of $\Dgt$ for various $\Te$ (assuming $\Te$=$\Trec$=$\Tadv$). Experimentally this means, given combinations of $\Te$, $\Dgt$ and  $\eta$ result in certain $\CaN$ i.e., the twin drops move at a certain speed. Theoretical (Fig.~\ref{fig:cadge}) and experimental (Fig.~\ref{fig:vNeck}) findings agree quantitatively (within $\approx\unit[10]{\%}$). Both show that $v_N$ is approximately independent from $\Dgt$ for a wide range of $\Dgt$ (which is quite counterintuitive, $\Dgt$ is obviously the source for the motion). The analysis predicts that the contact angles determine the order of magnitude for $\CaN$. 
Assuming experimentally realistic conditions, other parameters change $\CaN$ by less than $\pm25\%$. As depicted, changing $D$ by a factor of $100$ (e.g., due to shear-induced dispersion~[14]) barely affects $\CaN$. Different absolute or relative drop sizes also have little influence because the (large) neck curvature, not the (much smaller) drop curvatures dominates capillarity.

The analysis leading to Fig.~\ref{fig:cadge} assumes stationary conditions. This requires large drop volumes so that the flow between \Drec\ and \Dadv\  changes the composition of \Dadv\ and thus $\Dgt$ only slowly. Also, $\Dgt$ has to be sufficiently large so that minor changes in $\Dgt$ barely change $\CaN$ and $v_N$ is approximately constant (horizontal sections of $\CaN$ vs $\Dgt$ in Fig.~\ref{fig:cadge}). The range where $\CaN$ decreases rapidly with decreasing $\Dgt$ is quantitatively out of the scope of the analytical approach presented here. Experimentally, immediate coalescence is observed in this region. Nevertheless, via the flow through the neck, we can estimate with our approach the lifetime of the non-coalescence ~[14] (which agrees well with the experiments). 

\emph{Conclusion.---}We present new data from the state of non-coalescence of two sessile drops from different, completely miscible liquids and analyze the hydrodynamics of the (temporarily) stationary twin-drop configuration. In a thin film approximation the moving twin drop configuration is described as two moving drops that are connected by a liquid neck. Through this neck liquid flows from upstream to downstream drop. An advection-diffusion balance establishes a localized, (temporarily) stable surface tension gradient close to the neck, which causes a Marangoni flow that sucks liquid out of the neck. This counteracts the capillary-driven flow into the neck and thus stabilizes non-coalescence. The whole system forms a self-stabilizing, traveling wave (twin drop contour). The theoretical predictions are in quantitative agreement with the experimental findings. 

\acknowledgments{We thank H. M\"ohwald for scientific advice and general support and L. Pismen for helpful discussions. S.K. was supported by the DFG (RI529/16-1).}

%
%
%

%
\end{document}